\documentclass{elsart}
\usepackage{epsfig}
\usepackage{natbib}

\begin{document}

\begin{frontmatter}
\title{Probing the $sp^2$ dependence of elastic moduli in 
ultrahard diamond films}

\author{M. G. Fyta\thanksref{mfyta}}, 
\author{G. C. Hadjisavvas\thanksref{xsavvas}}, and
\author{P. C. Kelires}
\address{Physics Department, University of Crete, P.O. Box 2208, 
710 03, Heraklion, Crete, Greece}

\thanks[mfyta]{Present address: Department of Physics, Harvard University, 
17 Oxford Str., Cambridge, 02138 MA, USA.}
\thanks[xsavvas]{Present address: Department of Physics \& Astronomy,
 Vanderbilt University,
6631 Stevenson Center, Nashville, TN 37235, USA.}

\begin{abstract}
The structural and elastic properties of diamond nanocomposites and
ultrananocrystalline diamond films (UNCD) are investigated using both 
empirical potentials and tight binding schemes. We find that both
materials are extremely hard, but their superb diamondlike properties
are limited by their $sp^2$ component. In diamond composites, the 
$sp^2$ atoms are found in the matrix and far from the interface with the 
inclusion, and they are responsible for the softening of the material. 
In UNCD, the $sp^2$ atoms are located in the grain boundaries.
They offer relaxation mechanisms which relieve the strain but, on the
other hand, impose deformations that lead to softening.
The higher the $sp^2$ component the less rigid these materials are.
\end{abstract}


\end{frontmatter}

\section{Introduction}

Carbon is an extraordinary chemical element. 
Its ability to form a variety of hybridized bonds
leads to a plethora of structures with different properties. 
It is involved in most of the organic compounds in nature, but it is 
also the basic building block of many inorganic compounds and
materials which are important for technological purposes.
For example, diamondlike carbon structures have attracted intense research 
in both the fundamental and practical level. Pure carbon films that contain 
a high fraction of $sp^3$ diamondlike local geometries are potential 
candidates for incompressible and ultrahard materials. Examples of such 
materials, and the focus of the present study, are ultrananocrystalline 
diamond films and diamond nanocomposites. The realization of such materials 
is a promising route towards improving the mechanical properties 
of pure systems.

Ultrananocrystalline diamond (UNCD) films are about 1 $\mu$m thick and
show high crystallinity (for a review see Ref.\ \cite{Gruen}).
There seems to be only a small fraction of nondiamond phases 
(estimated below 5\%).
They possess a fairly uniform distribution of conglomerates made up of 
diamond crystallites with an average size between 3 and 5 nm. 
Crystallites larger than 10 nm are seen very occasionally, and these 
tend to be highly faulted and extensively twinned. 
Much work has been done, experimentally and theoretically, on the
 characterization of these films \cite{uncdCharact}.
 The main issues are the details
 of the structure and the nature of the grain 
boundaries (GB) \cite{bondingGB,modelGB}. These could be twisted, tilted,
 amorphous or just disordered boundaries. Nevertheless, as long as the
exact structure of the GB is not clear, their role in determining the 
properties of the film cannot be revealed. Regarding the atomistics,
due to the fact that the crystallite size is limited to the nanoscale,
the fraction of atoms located at the GB increases drastically with 
a decrease of the grain size.
The presence of $sp^2$ atoms, together with the experimental fact that
no graphitic phase was found between the GB suggests that
the atoms are $\pi$-bonded across sharply delineated GB
limited to widths of 0.2-0.5 nm.

Diamond nanocomposites ($n$D/$a$-C), on the other hand, 
consist of diamond crystallites
that are embedded in an amorphous carbon ($a$-C) matrix. 
The appropriate choice of the size of the nanodiamond inclusion
allows tailoring of the properties of the embedding medium.
It has been recently shown that diamonds embedded in a tetrahedral
$a$-C (t$a$-C) matrix are both stable and highly
incompressible materials, possessing elastic moduli higher than t$a$-C
\cite{nDfract}.
Their synthesis and growth mechanisms have been explored 
in both hydrogenated \cite{Lifshitz} and pure \cite{Yao} t$a$-C matrices. 

In the present work, we further explore some of the elastic properties
of both UNCD and $n$D/$a$-C, and try to correlate our findings to the
microstructure of these materials. We mainly focus on
the elastic parameters, such as the bulk and Young's moduli, 
which are evaluated at the equilibrium state and can 
provide information on the elastic response of the system to
compression and extension, respectively. Due to computational limits the 
grains modeled were restricted to sizes up to 3 nm. 
These sizes may be smaller than the average size seen in the 
experiments, but as it will be shown in the next sections,
the main features of the experimental findings are captured.

\section{Methodology and Structural Characteristics}

Here, we present results based on Monte Carlo (MC) simulations within
the Tersoff empirical potential (EP) approach \cite{Tersoff}.
This method provides a fairly good description of the structure and 
energetics of a wide range of $a$-C phases \cite{Kelires94,Kelires00},
and superb statistical accuracy. We also present results based on
Molecular Dynamics (MD) simulations, using the environment-dependent 
tight-binding (EDTB) model of Tang, Wang, Chan, and Ho \cite{TWCH},
which provides superior accuracy in the energetics compared to
the empirical schemes. This model
allows the tight-binding parameters to change according 
to the bonding environment and has been successfully used 
in recent applications \cite{Galli1,chrm_aCdens}.
The use of both MC/EP and MD/EDTM schemes allows the treatment
of large systems, and a check of the validity of the results
by comparison to more accurate ones. Large supercells with up to about
10000 atoms are treated with the MC/EP method, while the smaller ones
(less than 600 atoms) are modeled using the MD/EDTB scheme. Periodic
boundary conditions are imposed on all supercells.

The $n$D/$a$-C structures contain a spherical nanocrystal in the 
middle, surrounded by dense amorphous carbon. They have been 
constructed, as previously reported \cite{nDfract}, by applying 
the liquid-quench method to generate the amorphous
matrix and by subsequently relaxing all the atoms at low temperatures.
In the high density $n$D/$a$-C, where the nanodiamonds are 
embedded in a t$a$-C matrix ($sp^2$ fraction less than 30\%), the 
majority of the local geometries are $sp^3$ hybrids. It is essential in
understanding the properties of such structures to take a closer look at 
the few $sp^2$ atoms and investigate the link between $sp^2$ fraction
and nanodiamond size. We have previously found \cite{nDfract} that
$sp^2$ sites are located in the $a$-C matrix, 
far from the interface between the matrix and the embedded nanodiamond, 
and form small clusters and chains,  
similar to the case of pure t$a$-C networks, where the
clustering is also associated to stress relief \cite{Kelires00}.

\begin{figure}[h]
 \begin{center}
 \epsfig{file=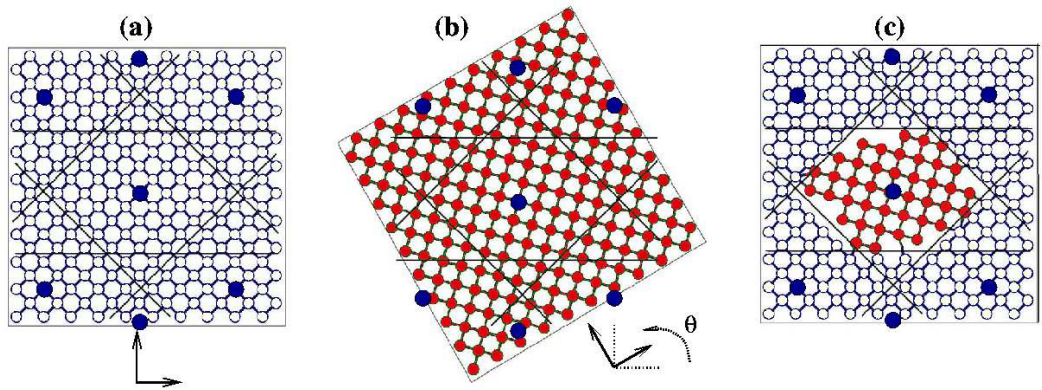, width=1\textwidth}
 \caption{Schematic representation 
of the construction of a nanodiamond cell. The large spheres
 denote the imaginary lattice points used to construct a WS cell
 by bisecting
 the lines which connect the central of these points to the others. 
 (a) The atoms inside a WS cell are removed.
 The lattice in panel (b) has been rotated by an angle $\theta$ with respect
 to panel (a). The atoms of the second WS cell are embedded in the 
respective region of the
 first cell. The generated structure, before relaxation, is shown in (c).}
 \label{fig1}
 \end{center}
 \end{figure}

The generation of the UNCD structures was done in a different manner. 
The concept of the Wigner-Seitz (WS) cell was used. 
The WS cell around a lattice point is defined as the set of points in space,
which are closer to that lattice point than any of the other lattice points.
The WS unit cell around a lattice point can be constructed by 
drawing lines connecting the point to all others in the lattice, 
bisecting each line with
a plane, and taking the smallest polyhedron containing the point bounded by 
these planes (Fig.\ref{fig1}a). In this way, starting from a cubic
diamond configuration, imaginary lattice points
are chosen in a  bcc (fcc) structure and a large WS cell is constructed 
having the shape of a truncated octahedron (rhombic dodecahedron). 
The atoms inside the WS cell are removed, leaving 
an empty space in the interior of the cubic lattice.
At a second step, another cubic diamond lattice is used, which is rotated
with respect to the first one (panel (b) in Fig.\ref{fig1}). 
The same WS cell, as in the first step, is constructed and the 
atomic configuration contained in this cell is subtracted and embedded
in the empty region of the first structure. In this way, misoriented
polyhedra, simulating the diamond nanograins are generated.
The grain sizes range from 0.7-3 nm. The final step is to fully relax
the structure, both the atomic positions and the volume, at 500-1000 K,
and then cool it to 0 K where properties are taken. The thin interface
region between the fully relaxed misoriented grains are the 
grain boundaries (GB).

Various random grain misorientations were tested during the generation of
the UNCD cells. The angles were in the range of about 5-45 degrees.
 The criterion for accepting or rejecting the
relaxed structures was the fraction of $sp^{2}$ sites. For a high percentage
 of threefold atoms (total number over 15-20 \%) the films were considered
as highly defective and deviated significantly from the experimental 
findings.
The larger the rotation angles, the higher the number of the 
threefold atoms.
This was actually expected, since for larger angles the mismatch
between the neighboring diamond grains becomes larger, giving rise
to more defects which are related to the threefold atoms. 
Another important issue is the grain size.
For similar grain misorientations, the larger grains result in fewer
threefold sites. Sizes below $\sim$ 1 nm in general lead to
unstable grains and very high disorder.

\begin{figure}[h]
 \begin{center}
 \epsfig{file=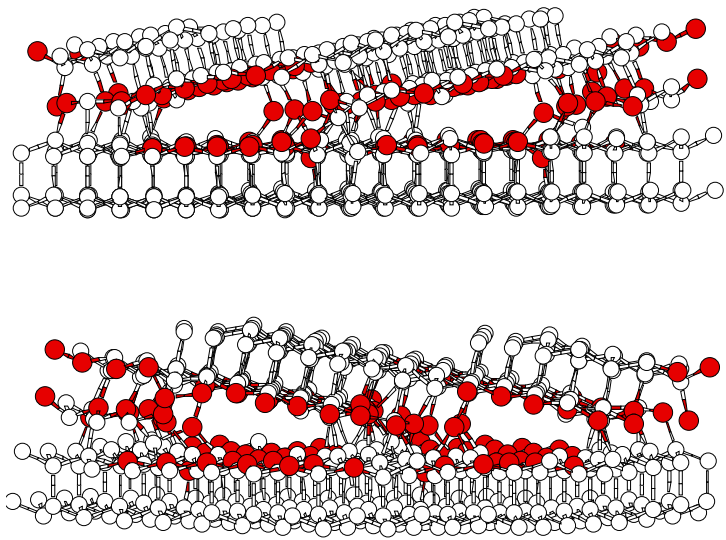, width=0.7 \textwidth}
 \caption{Side views of randomly oriented grain boundaries 
 in a nanodiamond cell.
 Open circles denote $sp^{3}$ hybridization, while the dark (red) ones are the
 $sp^{2}$ sites.}
 \label{fig2}
 \end{center}
 \end{figure}

Taking a deeper look into the GB reveals some
interesting features (Fig.\ref{fig2}). These are randomly oriented, 
since the grain misorientations were randomly chosen. Rotation angles that 
would lead to well defined and studied twist or tilted boundaries
were also chosen, but did not result in a relative low threefold fraction
\cite{bondingGB}, suggesting that in UNCD the grains are possibly
randomly oriented. The $sp^{2}$ sites are located only at the grain 
boundaries. The material therein is 
rather disordered than amorphous and the GB cannot be described 
as tilted or twisted. In most cases, the average  $sp^{2}$ component in 
the GB compared to the total number of atoms in the GB
 is $\sim$ 30 \% (similar to experimental findings). 
The width of the GB is approximately
3-3.5 \AA, again similar to experiment, and  deformations are
 actually restricted in this area.
 This, of course, varies depending on the grain misorientations.
For larger angles the width is larger, but the higher percentage of
threefold bonding is related to more deformed boundaries. 
Inspection of Fig.\ref{fig2} reveals an additional
feature. Small empty regions occur. The volume of these voids gets 
larger as the misorientations increase. The role of these voids is
not completely clear, but they could possibly be related to stress 
relief regions and assist in the softening of the material. The above
features are independent of the choice of the simulational method
used.

\section{Elastic Moduli and $sp^2$ component}

We now turn our attention to the elastic properties of the
networks. As representative quantities, we compute the bulk modulus $B$ 
and the Young's modulus $Y$. 
The former, is related to the response of
the system to a volume change, due to variations in pressure.
It is essentially a measure of a material's resistance to uniform compression.
A common procedure for the calculation of the bulk modulus, which
we also apply here uses the second derivative (with respect to 
the volume) of the Birch-Murnaghan isothermal equation of state \cite{murnachan}
Within this theory the energy is given as an analytical 
function of the volume to which we fit the
energy versus volume curves extracted from our simulations. 
These are produced by varying the volume for each structure,
allowing for relaxation in the atomic positions and 
calculating the corresponding energy.
As a complement to $B$, the Young's modulus serves
as a measure of the stiffness of a material.
It is defined as the rate of change of stress with strain 
(for small strains) and can easily be calculated through
the second derivative of the energy with respect to the applied strain.
For cubic systems, as the ones we deal with in this study, 
$Y$ is also efficiently given through $B$:
$Y=3B(1-2\nu)$, with $\nu=\frac{c_{12}}{c_{11}+c_{12}}$;
$c_{11},~c_{12}$ are elastic constants 
calculated using volume conserving orthorombic, additionally  
to monoclinic strains \cite{Kelires94}.
We refer to the cubic elastic moduli and accordingly derive $Y$,
based on the fact that the supercells we use are large, the strains 
are small and the deviation from cubic symmetry negligible.
The bulk modulus
$B$ for crystalline diamond calculated with ETDB and
the Tersoff potential is 428 GPa and 443 GPa, respectively. 
We also consider the Wooten-Winer-Weaire (WWW) model \cite{WWW}, which is a
fully tetrahedral structural model of $a$-C, i.e., a model of
``amorphous diamond'', whose
properties are an upper value for the corresponding
properties of single phase $a$-C and serve as a comparison to the properties
of the composites and UNCD. $B$ of WWW is 360 (365) GPa using the
Tersoff (EDTB) potential. Calculations of $Y$, or modulus of extension, 
with the Tersoff potential give values of 1060 GPa for diamond and 824 GPa 
for the WWW network \cite{Kelires94}.

We have recently shown using another tight-binding method that the 
elastic response of the
$n$D/$a$-C networks to hydrostatic deformation is controlled by the 
nanoinclusions \cite{nDfract}. 
Their bulk moduli are enhanced over the values for
single phase $a$-C. The enhancement is stronger for dense composites
(lower $sp^2$ component) and for larger diamond sizes.
Here, we extend that analysis by confirming these results
using different computational
schemes and also including some calculations of
the Young's moduli. The results are shown in Table \ref{tab1}, divided
in two parts. The left panel of the table represents the increase
of the moduli for composites that contain a 1.7 nm nanodiamond,
but differ in the $sp^2$ component of the matrix.
The right panel shows results for composites with similar high
density $a$-C matrices ($sp^2$ about 12\%), that differ in the
size of the nanodiamond they contain.
Although the variation in the diamond sizes used was not very large
it is clear from all simulations carried out and the range
of grain sizes modeled, that both $B$ and $Y$
are strongly affected by the $sp^2$ content rather than the size.

The results obtained using the EDTB method are similar. A composite
with a 1.05 nm diamond and matrix with a 30\% $sp^2$ fraction
has a bulk modulus of 300 GPa. The corresponding value for a 
pure $a$-C network with 
the same $sp^2$ fraction is about 12\% lower. The same composite with
a lower $sp^2$ component (11\%) in the matrix has an even higher bulk
modulus (357 GPa). Comparison with pure $a$-C reveals again that the 
modulus is about 11\% higher.
As the size of the diamond increases, the bulk modulus of the composite
increases and it is again enhanced compared to single-phase $a$-C. 
The modulus is 346 GPa for a 1.24 nm diamond embedded in an $a$-C matrix 
with 20\% threefold atoms, which is increased about
16\% over the $a$-C value.

Both schemes used, MC/EP and MD/EDTB, show very similar trends. The 
rigidity of the composites is considerably enhanced over that
of pure $a$-C, and this is attributed to the presence of the diamond
nanoinclusions. On the other hand, the increase of the $sp^2$ component 
in the matrix plays an important role in the lowering of the rigidity.
As previously mentioned, the $sp^2$ atoms are 
mainly located far from the interface with the inclusion and form
small chains and clusters. These clusters are known to be 
under tensile stress \cite{Kelires94}, providing relief of the
overall compressive stress of the material, but they also provide
softening mechanisms. Therefore, for an optimum performance of
nanodiamond composite films, a balanced choice of size and density
of nanoparticles and $sp^2$ content in the matrix is needed.

 \begin{table}[h]
 \begin{center}
 \caption{Bulk and Young's moduli (in GPa) for various 
$n$D/$a$-C composites.}
 \label{tab1}
 \begin{tabular}{|c|c|c|||c|c|c|}
\hline
\multicolumn{3}{|c|||}{D=1.7 nm} & \multicolumn{3}{|c|}{$sp^2$ (\%)=12}\\
\hline
 $sp^2$ (\%)&B&Y & D (nm)&B&Y \\ \hline
3.1&364 &830 &1.9&370&876\\ \hline
7.6& 344&  792 &2.0&374&884\\ \hline
20.5&241 &547  &2.2&375&901\\ \hline
 \end{tabular}
 \end{center}
 \end{table}

We next investigate the effect of the $sp^2$ component 
on the bulk and Young's moduli of various types of UNCD films.
These threefold atoms, located in the GBs, are believed to control
the elastic response of the films. The majority of the fourfold atoms 
in the grains are perfectly tetrahedral. A small fraction of $sp^3$
sites, located in the vicinity of the GBs are slightly distorted.
Therefore, they are not expected to play a significant
role in softening the material in comparison to crystalline diamond.
Also, as previously mentioned, the voids that occur in the GB, due to grain 
misorientations, could also assist in softening.
For small grain misorientations, these voids are not expected 
to greatly affect the properties, but should possibly play an
important role for large misorientations, in which cases
the softening is more evident.

\begin{figure}
 \begin{center}
 \epsfig{file=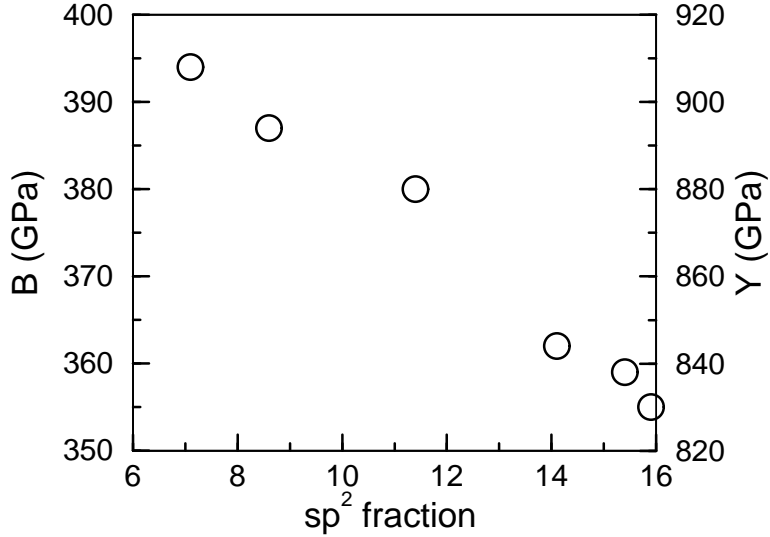, width=0.55\textwidth, angle=270}
 \caption{Bulk and Young's moduli for UNCD films
with grain sizes of 2.8 nm with varying $sp^{2}$ components (from
MC/EP calculations).}
 \label{fig3}
 \end{center}
 \end{figure}

The bulk and Young's moduli for UNCD structures with a fixed
grain size (2.8 nm) and different grain orientations, leading to
various $sp^{2}$ contents in the GBs, are plotted 
in Fig.\ref{fig3}. All moduli are lower than diamond's but still very high,
justifying the consideration of UNCD films as candidates for ultra-hard 
MEMS/NEMS components. The moduli are higher than those 
of pure $a$-C with similar threefold component, as well as those of the 
WWW ``amorphous diamond'' network. UNCD structures with low $sp^{2}$
component possess Young's moduli over 900 GPa.
The reported experimental values are in the range of 840-950 GPa, 
in very good agreement with the present theoretical predictions.
It is clearly shown in this graph, that the $sp^{2}$ atoms formed
due to the grain misorientations control the moduli.
A structure that consists of 2.8 nm grains has a
$sp^{2}$ fraction of 10\% and a bulk modulus of 379 GPa, while 
values for 1.4 nm grains are 13\% and 365 GPa, respectively. 
The grain misorientations were randomly chosen. The aim
was not to simulate specific grain boundaries, but to investigate how the
random misorientations determine the $sp^2$ fraction and thereby the
properties of the material.
In this way, not all the possible types of GB have been examined,
and not the whole region of the experimentally produced grain 
sizes was spanned. Nevertheless, the good comparison to experimental data
indicates that the main features related to the $sp^2$ content in 
UNCD films are well captured, and this work can serve as a first
step to further explore these films.

The connection of the moduli to the $sp^{2}$ sites in the GBs is also 
confirmed through the EDTB calculations, although smaller grain
sizes ($\sim$ 0.7-1.4 nm) were simulated in this case. In addition,
the dependence on the grain size was explored. For example,  
a UNCD film that consists of 1.4 nm grains with a 
misorientation around 15 degrees, has a threefold component $\sim$ 
8 \% and a bulk modulus of 310 GPa. This value is considerably
lower than the value of over 380 GPa computed for a nanograin size
of 2.8 nm in Fig. \ref{fig3}. The large difference is attributed to the small
size of the grains which becomes comparable to the GB size. Thus,
our simulations (both MC/EP and MD/EDTB) show that 
not only the grain misorientations have to be small, but also
the grains should be larger than 2 nm. We believe that model UNCD
structures with such characteristics (low $sp^{2}$ component, high
phase purity and high elastic moduli) realistically simulate
experimental UNCD films 

Finally, we compare the elastic moduli of similar, in terms of their 
$sp^{2}$ component, $n$D/$a$-C and UNCD films.
At low $sp^2$ components, the moduli of UNCD structures are
undoubtedly higher than those of $n$D/$a$-C composites. However,
the moduli of UNCD decrease more abruptly as $sp^{2}$ content increases
(see variation in Fig. \ref{fig3}). Thus, it seems that for optimum UNCD
films, the $sp^{2}$ content in the GBs need to be less than 10 \%.
Further calculations are required to unambigiously establish the link
between $sp^{2}$ component and GB structure. This includes the
consideration of a wider range of GB types and grain sizes.
From a technological point of view, our results
prove expedient in the quest for novel carbon materials
and serve as a guide toward engineering diamond films with desired
properties. Specifically, we provide a pathway to manipulate
the mechanical properties of crystalline diamond by efficiently 
choosing the grain size and controlling the $sp^2$ content in the structure.

\section{Conclusions}

We have studied in this work the link of the microstructure,
especially the $sp^2$ component, to the elastic moduli 
of diamond nanocomposites and ultrananocrystalline diamond films,
by employing complementary MC/EP and MD/EDTB approaches.
Both materials are found to be ultrahard, with moduli approaching
those of diamond. UNCD films are superior when having low $sp^{2}$ 
component in the GBs. On the other hand, $sp^{2}$ sites are
strain-relieving geometries in the GBs, so optimum values of
$sp^{2}$ content are required. This depends on grain size and
misorientation. For the same reason, a balanced choice of the $sp^{2}$
component is also needed in the case of diamond nanocomposites.
Further work has to be devoted for the elucidation of this issue.

{\bf Acknowledgments} 

The authors wish to thank G. Kopidakis for providing access to his TBMD 
code. The work was supported by a grant from the EU and the Ministry of 
National Education and Religious Affairs of Greece through the action 
``E$\Pi$EAEK'' (programme ``$\Pi\Upsilon\Theta$A$\Gamma$OPA$\Sigma$''.)

\end{document}